\documentclass[final,3p,times,twocolumn]{elsarticle}

\usepackage{lineno}

\usepackage{amsmath, amssymb}
\usepackage{graphicx}
\usepackage{dcolumn}
\usepackage{bm}

\usepackage[english]{babel}
\usepackage[utf8x]{inputenc}
\usepackage[T1]{fontenc}
\usepackage[colorinlistoftodos]{todonotes}
\usepackage{braket}
\DeclareMathAlphabet{\mathpzc}{OT1}{pzc}{m}{it}
\usepackage{titlesec}
\usepackage{babel,blindtext}
\usepackage{mathrsfs}
\usepackage{gensymb}
\usepackage{sidecap}
\usepackage{float}

\usepackage{ctable}
\usepackage{appendix}
\usepackage{subcaption}
\usepackage{ragged2e}
\usepackage{mathtools}
\usepackage{natbib}
\usepackage{subcaption}
\usepackage{ragged2e}
\usepackage{bm}
\DeclareCaptionJustification{justified}{\justifying}
\usepackage[a4paper=true,pagebackref=true]{hyperref}
\hypersetup{colorlinks = true, linkcolor = green, anchorcolor = red, citecolor = blue, filecolor = red, pagecolor = red, urlcolor = red}

\usepackage{cuted}
\usepackage{flushend}
\usepackage{widetext}
\usepackage{multicol}
\usepackage{pythontex}
\biboptions{sort&compress}

\begin{document}

\begin{frontmatter}


\title{Mixed state geometric phase for neutrino oscillations}



\author[1,2]{Sandeep Joshi}

\address[1]{Nuclear Physics Division, Bhabha Atomic Research Centre, Mumbai 400085, India }

\address[2]{Homi Bhabha National Institute, Anushakti Nagar, Mumbai 400094, India}

\ead{sjoshi@barc.gov.in}

\begin{abstract}
The geometric picture of neutrino oscillations offers a unique way to study the quantum mechanics of this phenomenon. In this picture, the propagation of a neutrino beam is described by a density matrix evolving in a state space with non-trivial geometry. We derive explicit expressions of the mixed state geometric phase which arise during such an evolution for both two and three flavor neutrino oscillations.  We show that, in the case of two flavor neutrino oscillations, the geometric phase is independent of the Majorana phase and it can be used as a measure of coherence of the neutrino beam.  
\end{abstract}




\end{frontmatter}


\section{Introduction}
The success of the theory of neutrino oscillations has led to many studies exploring the intricacies associated with this phenomenon. In the standard plane wave treatment, the neutrino flavor oscillations arise due to mixing and interference between massive neutrino states. A pictorial way to represent neutrino oscillations is in terms of precession of spin-polarization vector in presence of an effective magnetic field \cite{Kim:1987bv, Kim:1987ss}.  In particular, for the case of two-flavor oscillations, the polarization vector becomes three-dimensional and its precession can be easily visualized in both constant and time varying magnetic fields, the magnitude and direction of the magnetic field being specified by the Hamiltonian governing neutrino propagation. 

Quantum mechanically, such a precession can be understood in terms of evolution of the state vector in the system's Hilbert space. Such an evolution in the state space with non-trivial geometry gives rise to geometric phase. Let $\mathcal{H}$ denote the Hilbert space and $\mathcal{N}$ denote the set of normalized states in $\mathcal{H}$. The two vectors $\ket{\psi_1},\ket{\psi_2} \in \mathcal{N}$ represent the same physical state if $\ket{\psi_2}= e^{i \phi} \ket{\psi_1}$ where $\phi$ is real. The set of physical states is called the projective Hilbert space and is the ray space: $\mathcal{P}= \mathcal{N}/U(1)$. If $\mathcal{H}$ has complex dimension $n$, then $\mathcal{P}$ is a complex projective space of dimension $n-1$, $\mathcal{P}= \mathbb{C}P^{n-1}$. The projection map $\pi: \mathcal{N} \rightarrow \mathcal{P}$ maps each vector in $\mathcal{N}$ to its corresponding ray. The above construction defines the principle fibre bundle picture of the state space \cite{Simon:1983mh}. The bundle space $\mathcal{N}$ consists of three parts: the base manifold $ \mathcal{P}$, the fibre which is group $U(1)$ element attached to each point of the base manifold, and the map $\pi:\mathcal{N} \rightarrow \mathcal{P}$. Now consider the evolution of a normalized state $\ket{\psi(t)}: t\in [0, \tau]$ . Let the evolution is governed by Schr\"{o}dinger equation, so that the unitary evolution $\ket{\psi(0)}\rightarrow \ket{\psi(t)}= \mathcal{U}(t) \ket{\psi(0)}$ traces a curve $\mathcal{C}$ in $\mathcal{N}$. The projection $\pi: \ket{\psi(t)} \rightarrow \ket{\psi(t)} \bra{\psi(t)}$ gives the corresponding curve $\pi(\mathcal{C})= \mathscr{C}$ in $\mathcal{P}$. The evolution is cyclic if the curve $\mathscr{C}$ in closed i.e. $\ket{\psi(\tau)} \bra{\psi(\tau)}= \ket{\psi(0)} \bra{\psi(0)}$. In this case the corresponding curve $\mathcal{C}$ in the bundle space begins and ends on the same fibre such that $\ket{\psi(\tau)} = e^{i \phi_T} \ket{\psi(0)}$, where $\phi_T$ is the total phase acquired by the state during cyclic evolution. Now, there can be infinitely many curves in $\mathcal{N}$ which project to a given closed curve $\mathscr{C}$ in $\mathcal{P}$. It was shown in \cite{Aharonov:1987gg} that given a curve $\mathscr{C}$ in $\mathcal{P}$, we can define a functional of $\mathscr{C}$ called geometric phase which is independent of $\phi_T$ and the curve $C$ in the bundle space. The geometric phase is simply obtained by subtracting the dynamical phase from the total phase:
\begin{align}\label{eq1}
    \phi_G=& \arg\{\braket{\psi(0)| \mathcal{U}(\tau)| \psi(0}\} \nonumber \\+& i \int_0^\tau dt \braket{\psi(0)|\mathcal{U}^{\dagger}(t)\mathcal{\Dot{U}}(t)|\psi(0) }.
\end{align}
It can be shown that $\phi_G$ defined above is $(i)$ gauge invariant i.e. invariant under local phase transformations of $\ket{\psi}$ and $(ii)$ reparametrization invariant i.e. independent of parameter $t$ of $\mathcal{C}$. Thus $\phi_G$ is independent of the dynamics of $\ket{\psi(t)}$  and is a geometric property of the curve $\mathscr{C}$ in $\mathcal{P}$. Also since $\ket{\psi(t)}$ need not be an eigenstate of the Hamiltonian $H(t)$, hence the condition of adiabaticity and cyclicity of $H(t)$ are not required. Thus \eqref{eq1} generalizes the adiabatic Berry phase \cite{Berry:1984jv} to non-adiabatic situations. The definition of geometric phase has further been generalized to include non-cyclic and non-unitary evolution \cite{Samuel:1988zz, aitchison1992real, Mukunda:1991rc, pati1995geometric}, which has found numerous applications in physics \cite{Mead:1992zz, Jain:1998zz}.

Returning to two flavor neutrino oscillations, the Hilbert space in this case is the two dimensional complex space $\mathcal{H}= \mathbb{C}^2$. The space of normalized states is the unit sphere $\mathcal{N} = S^3$. Thus the projective Hilbert space is the complex projective line $CP^{1}= S^3/U(1)$ which is the Bloch Sphere $S^2$. The pure neutrino states correspond to points on the surface of the Bloch sphere. For the case of neutrino oscillations in vacuum or in a medium with constant density, the cyclic evolution of neutrino eigenstates produces a closed curve on $S^2$. The resulting geometric phase is equal to the standard expression of one half the solid angle subtended by the closed curve at the centre of the sphere \cite{Joshi:2017vpi}.

In the context of neutrinos various authors have derived explicit expressions of geometric phase in different settings, for example, neutrino oscillations in vacuum \cite{Blasone:1999tq, Wang:2000ep, Law:2007fb, Mehta:2009ea}, neutrino oscillations in medium with or without dissipation \cite{Naumov:1991ju,Naumov:1993vz,He:2004zc, Dajka:2011zz,Capolupo:2016idi,Wang:2015tqp,Dixit:2017ron}, neutrino spin-flavor oscillations \cite{Smirnov:1991ia, Vidal:1990fr, Joshi:2016unj, Joshi:2017vpi} and neutrino self-interactions \cite{Johns:2016wjd}. In all of the above cases the neutrino eigenstate undergoing evolution is considered as a pure state which can be expressed as a coherent superposition of different neutrino states. However, it has been shown that a neutrino produced in a charged-current interaction cannot be described by a pure state \cite{Giunti:2010ec}. The neutrinos produced in such a process are described by an incoherent superposition which is essentially a mixed state. 

In this present work we calculate the mixed state geometric phase for the case of neutrino oscillations in vacuum using the gauge invariant formulation \cite{singh2003geometric}. We show that our expression of the mixed-state geometric phase generalizes the previously obtained expressions by various authors for both two and three flavor neutrino oscillations. In Section 2, we describe the mixed state geometric phase for unitary evolution. In Section 3, we consider the two flavor case and derive explicit expression of mixed state geometric phase. We also compare the obtained expression of geometric phase with that of quantum coherence. In Section 4, we extend our calculation to three flavor neutrino oscillations and finally conclude in Section 5. 
\section{Mixed state geometric phase}
The mixed states are mathematically represented by density matrices which are convex sum of pure states projection operators. The notion of geometric phase for mixed state was first proposed by Uhlmann \cite{uhlmann1986parallel} using a procedure known as purification, in which the mixed state density matrix of the system is written as partial trace of a pure state density matrix of an extended system consisting of the given system and an ancilla.  An alternative definition of the mixed state geometric phase is given by Sj\"{o}qvist \textit{et. al.}  \cite{sjoqvist2000geometric} which is a direct generalization of the pure state geometric phase. While Ulhmann's formulation of geometric phase is based on purely mathematical ground, the definition by Sj\"{o}qvist \textit{et. al.} has a physical interpretation in the context of quantum interferometry \cite{sjoqvist2000geometric,pati2003geometric, chruscinski2012geometric, sjoqvist2015geometric}.  For a given unitary evolution, the above two approaches in general yield different results for the mixed state geometric phase. However, both of them reduce to the same expression for the case of pure states \cite{slater2002mixed, ericsson2003mixed}. In this paper we follow the approach formulated by Sj\"{o}qvist \textit{et. al.} and its subsequent gauge invariant generalization \cite{singh2003geometric}, since its physical implications are more transparent in the context of neutrino oscillations. 

Consider a mixed state density matrix undergoing a unitary evolution $\rho(0) \rightarrow \rho(t)= \mathcal{U}(t) \rho(0) \mathcal{U}^{\dagger}(t) $ which produces a curve $\Gamma:t\in[0, \tau]$ in the space of density operators. Let initial density matrix has the diagonal form 
\begin{equation} \label{eq2}
    \rho(0)= \sum_{k=1}^N w_k \ket{k}\bra{k} ,
\end{equation}
where $N$ is the dimension of the Hilbert space. Then unitarily evolved density matrix can be expressed as   
\begin{equation} \label{eq3}
    \rho(t)= \sum_{k=1}^N w_k \ket{k(t)}\bra{k(t)}  ,
\end{equation} 
where $\ket{k(t)}= \mathcal{U}(t)\ket{k}$. The phase shift acquired by $\rho(t)$ relative to $\rho(0)$ is given by  \cite{sjoqvist2000geometric} 
\begin{equation} \label{eq4}
    \gamma_T= \arg \bigg\{\mbox{Tr}\big[\mathcal{U}(\tau)\rho(0)\big]\bigg\} = \arg \bigg\{\sum_{k=1}^N w_k \braket{k| k(\tau)}\bigg\}. 
\end{equation}
The above formula can be verified by analyzing the interference pattern in a Mach-Zehnder interferometer where the input beam is the mixed state \eqref{eq2}. After splitting the beam, one arm of the interferometer is exposed to a variable $U(1)$ phase shift $e^{i \chi}$ and the other arm to the unitary operator $\mathcal{U}(t)$. On recombining the two beams, the output intensity shows the following interference  profile \cite{sjoqvist2000geometric}: 
\begin{equation}
    I = 2\big(1+ |\mbox{Tr}\big[\mathcal{U}(t)\rho(0)\big]| \cos\big(\chi- \arg \mbox{Tr}\big[ \mathcal{U}(t)\rho(0)\big]\big)\big).
\end{equation}
The above interference pattern clearly shows that \eqref{eq4} correctly describes the relative phase shifts between $\rho(0)$ and $\rho(t)$. In addition, several experimental tests have confirmed the validity of \eqref{eq4} (see \cite{sjoqvist2015geometric} for references).

The dynamical phase for the mixed state can be defined as the time integral of the average of Hamiltonian $H(t)$
\begin{align} \label{eq6}
    \gamma_D=& - \int_0^\tau dt \mbox{Tr} \big[\rho(t)H(t)\big] \nonumber \\ =& - i \int_0^{\tau} dt \mbox{Tr} \big[\rho(0)\mathcal{U}(t)^\dagger \mathcal{\dot{U}}(t)\big].  
\end{align}
The geometric phase in this case, however, cannot be simply obtained by subtracting accumulated phase $\eqref{eq6}$ from the total relative phase \eqref{eq4} due to the weight factors appearing in the two terms. To circumvent the issue, one defines the notion of parallel transport in which the dynamical phase \eqref{eq6} vanishes identically and thus the phase acquired by the mixed state during evolution is purely geometric. This can  be  done by requiring $\rho(t)$ and $\rho(t+dt)$ to be \textit{in phase}, which leads to the condition \cite{sjoqvist2000geometric}
\begin{equation}\label{eq7}
    \mbox{Tr} \big[\rho(t)  \mathcal{\dot{U}}(t) \mathcal{U}(t)^\dagger  \big]= 0.
\end{equation}
However, the condition \eqref{eq7} is not sufficient and a stronger condition is required, in which each eigenstate of the density matrix is parallel transported \cite{sjoqvist2000geometric}:
\begin{equation} \label{eq8}
    \braket{k|\mathcal{U}(t)^\dagger \mathcal{\dot{U}}(t)|k}= 0,~ k=1,2,...,N.
\end{equation}
It has been shown that one can incorporate the above conditions in a gauge invariant functional which depends only on the curve $\Gamma$ and has the following form \cite{singh2003geometric}: 
\begin{equation} \label{eq9}
    \gamma_G= \arg \bigg\{ \sum_k w_k \braket{k|k(\tau)} \exp\bigg(-\int_0^{\tau} dt \braket{k(t)|\dot{k}(t)}\bigg)\bigg\}.
\end{equation}
It can be seen that imposing the parallel transport conditions \eqref{eq8}, the above expression reduces to the total phase \eqref{eq4}. Also, for the case of pure states undergoing cyclic evolution, \eqref{eq9} reduces to the geometric phase \eqref{eq1}. Thus \eqref{eq9} provides a gauge invariant expression for the mixed state geometric phase.   
\section{Two flavor neutrino oscillations}
The case of two flavor neutrino oscillations gives us a useful toy model to study the important quantum mechanical features of the phenomenon. In this case the space of mixed states is the unit ball in $\mathbb{R}^3$, also called as Bloch ball. Pure neutrino states lie on the extremal points of the Bloch ball, which correspond to the Bloch sphere $S^2$. Thus a general mixed state can be represented as a point in the interior of the Bloch sphere. As the neutrinos undergo flavor oscillations, the unitary evolution of the mixed state traces a curve on a spherical shell with radius equal to length of the initial polarization vector. Due to non-trivial geometry of the underlying state space  the above curve gives rise to geometric phase which can be calculated using \eqref{eq9}.  

To this end, we consider a beam of neutrinos characterized by the following density matrix in the flavor basis
\begin{equation} \label{eq10}
    \hat{\rho} = \sum_{\alpha= e, \mu} w_\alpha \ket{\nu_\alpha} \bra{\nu_\alpha},
\end{equation}
where $w_\alpha$ is the initial statistical weight of the flavor state $\ket{\nu_\alpha}$, such that $\sum_\alpha w_\alpha = 1$. The density matrix \eqref{eq10} describes an incoherent mixture of different neutrino flavors, which are generated in a single or multiple weak interaction processes\cite{Giunti:2010ec, Giunti:2007ry}.  The flavor states are related to mass eigenstates via unitary transformation 
\begin{equation} \label{eq11}
    \ket{\nu_\alpha}= \sum_{i=1,2} U^{\ast}_{\alpha i} \ket{\nu_i},
\end{equation}
where $U$ is called mixing matrix. For vacuum oscillations, $\ket{\nu_i}$ are the eigenstates of the propagation Hamiltonian with energy eigenvalue $E_i= \sqrt{p_i^2+ m_i^2}$, where $p_i$ and $m_i$ represent the momentum and mass of the $i$th mass eigenstate. The mixing matrix, for the case of two flavor oscillations in vacuum, can be expressed as
\begin{equation} \label{eq12}
    U = \begin{pmatrix}
    \cos \theta & e^{i \phi} \sin \theta \\
    - e^{-i \phi} \sin \theta & \cos \theta
    \end{pmatrix},
\end{equation}
where $\theta$ is the vacuum mixing angle and $\phi$ is the Majorana phase. In the standard plane wave approximation, the Schr\"{o}dinger evolution of the mass eigenstates is given by 
\begin{equation}\label{eq13}
    \ket{\nu_i(x,t)} = e^{- i E_i t+ i p_i x} \ket{\nu_i},
\end{equation}
where the space-time interval $(x,t)$ is the separation between the propagation and production point, and we have written $\ket{\nu_i(0, 0)}$ as $\ket{\nu_i}$ for brevity. For the case of \textit{ultra-relativistic} neutrinos, one can employ the approximation $x \approx t$, under which \eqref{eq13} becomes 
\begin{equation}
     \ket{\nu_i(t)} = e^{- i m_i^2 t/2 E} \ket{\nu_i},
\end{equation}
where $E$ represents the neutrino energy obtained after neglecting the  mass contributions. Thus the time evolution of the flavor states \eqref{eq11} can be written as
\begin{equation} \label{eq15}
     \ket{\nu_\alpha(t)}=  \sum_i U^{\ast}_{\alpha i} e^{-i m_i^2 t/2 E } \ket{\nu_i}.
\end{equation}
The amplitude of  $\nu_\alpha \rightarrow \nu_\beta$ transition can now be obtained using \eqref{eq11} and \eqref{eq15}
\begin{equation} \label{eq16}
    \psi_{\alpha \beta}(t) = \braket{\nu_\beta| \nu_\alpha(t)}= \sum_i U^{\ast}_{\alpha i} U_{\beta i} e^{-i m_i^2 t/2 E}.
\end{equation}
The initial state of the neutrino beam is described by  the density matrix 
\begin{equation} \label{eq17}
    \rho(0) = \begin{pmatrix}
   w_e & 0 \\
    0 & w_\mu
    \end{pmatrix}.
\end{equation}
As the beam propagates in space, the state undergoes a unitary evolution $\rho(t)= \mathcal{U}(t) \rho(0) \mathcal{U}(t)^{\dagger}$, where $\mathcal{U}(t)$ is the unitary evolution operator given by 
\begin{equation}\label{eq18}
    \mathcal{U}(t) = \begin{pmatrix}
    \psi_{ee}(t) & \psi_{\mu e}(t) \\
    \psi_{e \mu}(t) & \psi_{\mu \mu}(t)
    \end{pmatrix}.
\end{equation}
The density matrix at time $t$ we can be written using \eqref{eq17} and \eqref{eq18} 
\begin{equation} \label{eq19}
    \rho(t)= \begin{pmatrix}
   w_e|\psi_{ee}(t)|^2+ w_\mu |\psi_{\mu e}(t)|^2 & (w_e- w_\mu)\psi_{ee}(t) \psi_{e\mu}^{\ast}(t) \\
   (w_e- w_\mu)\psi_{ee}^{\ast}(t) \psi_{e\mu}(t)  & w_e|\psi_{e\mu}(t)|^2+ w_\mu |\psi_{\mu \mu}(t)|^2
    \end{pmatrix},
\end{equation}
where we have used the unitarity relation 
\begin{equation}\label{eq20}
    \psi_{ee} \psi_{e \mu}^{\ast}=- \psi_{\mu e} \psi_{\mu \mu}^{\ast}.
\end{equation}
The explicit form of the transition amplitudes can be obtained using \eqref{eq12} and \eqref{eq16} : 
\begin{align} \label{eq21}
    \psi_{ee}(t) =& e^{i \omega_p t/2}\cos^2 \theta+ e^{-i \omega_p t/2} \sin^2 \theta , \nonumber \\
    \psi_{e \mu }(t) =& - e^{-i \phi} (e^{i \omega_p t/2}- e^{-i \omega_p t/2})\sin \theta \cos \theta ,\nonumber \\
     \psi_{\mu e }(t) =& - e^{i \phi} (e^{i \omega_p t/2}- e^{-i \omega_p t/2})\sin \theta \cos \theta ,\nonumber \\
      \psi_{\mu \mu}(t) =& e^{i \omega_p t/2}\sin^2 \theta+ e^{-i \omega_p t/2} \cos^2 \theta ,
\end{align}
where $\omega_p= \Delta m^2/2E$, $\Delta m^2 = m_2^2-m_1^2$ being the mass-squared difference. $\omega_p$ can be physically interpreted as a precession frequency. To see this, consider the neutrino Hamiltonian in the flavor basis
\begin{equation}
\label{eq22}
H_{f}= \frac{\Delta m^{2}}{4 E}
\begin{pmatrix}
-\cos 2 \theta & e^{i \phi}\sin 2 \theta
\\
e^{-i \phi}\sin 2 \theta & \cos 2 \theta
\end{pmatrix}
= \frac{\omega _{p}}{2} \textbf{B}\cdot \bm{\sigma },
\end{equation}
where $\textbf{B}= (\sin 2 \theta \cos \phi, -\sin \theta \sin \phi,-\cos 2 \theta )$  are the Pauli matrices. 
An equivalent way to express \eqref{eq21} is in the following form \cite{fano1957description}:
\begin{equation}\label{eq23}
    \rho(t)= \frac{1}{2}\big(1+ {\bf{P}}(t)\cdot\bm{\sigma}\big),
\end{equation}
where $\bf{P}$ $= (P_x,P_y,P_z)$ is called polarization vector. The evolution of density matrix is given by von Neumann equation:
\begin{equation} \label{eq24}
    i \frac{d \rho}{dt} = \big[H_f, \rho\big].
\end{equation}
For the given Hamiltonian \eqref{eq22} and density matrix \eqref{eq23}, we obtain
\begin{equation}\label{eq25}
    \frac{d {\bf{P}}}{dt}= \omega_p \big(\bf{B}\times {\bf{P}} \big).
\end{equation}
The geometric interpretation of neutrino oscillations can now be clearly seen from \eqref{eq22} and \eqref{eq25}. Specifically, \eqref{eq25} represents the precession of polarization vector $\bf{P}$ around a magnetic field $\bf{B}$, with precession frequency $\omega_p$. The initial value of polarization vector can be obtained by comparing \eqref{eq17} and \eqref{eq23}, ${\bf{P}}(0)= (0, 0, w_e-w_\mu)$. For the case of pure neutrino states $w_e= 1(0)$ and $w_\mu = 0(1)$. In this case we have $P_z = \pm 1$. Thus the neutrino states correspond to points on the unit sphere $S^2$, $\ket{\nu_e}$ and $\ket{\nu_\mu}$ being the antipodal points. In addition, \eqref{eq25} shows that for constant $\bf{B}$, the length of the polarization vector remains unchanged. Thus the precession of the polarization vector will trace a curve on the Bloch sphere. The geometric phase associated with this curve has been calculated for both cyclic \cite{Blasone:1999tq} and non-cyclic \cite{Wang:2000ep} cases. 

Now, the general incoherent mixture of neutrino flavor states is described by polarization vector with length less than unity. However, \eqref{eq25} still remains applicable, which implies that during precession the initial length, given by $P_z(0)= w_e- w_\mu$, remains unchanged. The precession of the component $P_z(t)$, which is related to the transition probabilities, can be obtained by comparing \eqref{eq19} and \eqref{eq23}
\begin{equation}
    P_z(t)= \big(w_e- w_\mu\big) \big(-1+ 2 |\psi_{ee}(t)|^2\big).
\end{equation}
The precession equations for $P_x(t)$ and $P_y(t)$ can be obtained in a similar manner. Geometrically, the precession can be visualized as being described by a  cone of length $P_z(0)$ with axis along $\bf{B}$ and opening angle $2 \theta$. Thus such a  precession will trace a curve $\Gamma$ on a spherical shell with radius $P_z(0)$. To evaluate geometric phase in this case, we first note that the initial density matrix \eqref{eq17} is diagonal, so its eigenvectors are simply given by :
\begin{equation}
    \ket{e_1} = \begin{pmatrix}
   1 \\
     0
    \end{pmatrix}, ~
      \ket{e_2} = \begin{pmatrix}
    0 \\
    1
    \end{pmatrix}.
\end{equation}
The eigenvectors of density matrix \eqref{eq19} can now be obtained using $\ket{e_i(t)} = \mathcal{U}(t) \ket{e_i}; i=1,2$, which gives
\begin{equation}
    \ket{e_1(t)} = \begin{pmatrix}
    \psi_{ee}(t) \\
     \psi_{e \mu}(t)
    \end{pmatrix}, ~
      \ket{e_2(t)} = \begin{pmatrix}
    \psi_{\mu e}(t) \\
     \psi_{\mu \mu}(t)
    \end{pmatrix}.
\end{equation}
 Now using expression \eqref{eq9}, we obtain the following form for the geometric phase: 
\begin{align} \label{eq29}
    \gamma_G =& \arg \Big\{w_e \psi_{ee}(\tau) \exp \Big(-\int_0^\tau dt \big(\psi_{ee}^{\ast}(t) \Dot{\psi}_{ee}(t)+\psi_{e\mu}^{\ast}(t) \Dot{\psi}_{e\mu}(t)\big) \Big) \nonumber \\ &
    + w_\mu \psi_{\mu \mu}(\tau) \exp \Big(-\int_0^\tau dt \big(\psi_{\mu e}^{\ast}(t) \Dot{\psi}_{\mu e}(t)+\psi_{\mu \mu}^{\ast}(t) \Dot{\psi}_{\mu \mu}(t)\big)\Big)\Big\}.
\end{align}
Substituting the explicit values of probability amplitudes from \eqref{eq21}, we obtain
\begin{align} \label{eq30}
    \gamma_G= \arg \Big\{w_e \big(e^{i \omega_p \tau/2}\cos^2 \theta + e^{-i \omega_p \tau/2} \sin^2 \theta \big) e^{-i \omega_p \tau \cos 2 \theta/2 } \nonumber \\ 
    + w_\mu \big(e^{i \omega_p \tau/2}\sin^2 \theta + e^{-i \omega_p \tau/2} \cos^2 \theta \big) e^{i \omega_p \tau \cos 2 \theta/2 })\Big\}.
\end{align}
Finally after rearranging the terms, we can write this equation as
\begin{widetext}
\begin{equation} \label{eq31}
    \gamma_G= \tan^{-1} \Bigg(\frac{(w_e - w_\mu) \Big(- \tan\big(\omega_p \tau \cos 2 \theta/2\big) + \cos 2 \theta \tan \big(\omega_p \tau /2\big)\Big)}{1+ \cos 2 \theta\tan \big(\omega_p \tau /2\big) \tan\big(\omega_p \tau \cos 2 \theta/2\big) } \Bigg).
\end{equation}
\end{widetext}
The above expression constitutes the central result of the paper. An important point to observe in \eqref{eq31} is that it is independent of the Majorana phase $\phi$. Thus the mixed state geometric phase for two flavor neutrino oscillations does not distinguish between Dirac and Majorana neutrinos. Since the geometric phase depends only on the curve $\Gamma$, during the evolution both Dirac and Majorana  neutrino flavor states trace the same curve in the space of density operators, despite having different evolutions in the Hilbert space. 

It can be shown that for pure neutrino states the geometric phase \eqref{eq31} reduces to earlier obtained results by various authors. \\
$(i)$ \textit{Noncyclic geometric phase}.  Consider the evolution of state $\ket{\nu_e}$, for which $w_e =1$ and $w_\mu = 0$. Substituting these weight factors in \eqref{eq31}, we obtain 
\begin{equation} \label{eq32}
    \gamma_G^P = - \frac{\omega_p \tau}{2} \cos 2 \theta + \tan^{-1} \Big(\cos 2 \theta \tan \frac{\omega_p \tau }{2}\Big).
\end{equation}
This is the noncyclic geometric phase for the pure flavor state $\ket{\nu_e}$ as obtained in Ref. \cite{Wang:2000ep}. \\
$(ii)$ \textit{Aharonov-Anandan phase}. Let us now consider the cyclic evolution of the mixed state which corresponds to $\tau= 2 \pi/\omega_p$. In this case \eqref{eq31} becomes 
\begin{equation} \label{eq33}
    \gamma_G^{AA}= \tan^{-1}\Big(\big(w_e- w_{\mu}\big) \tan\big(\Omega/2\big)\Big),
\end{equation}
where $\Omega= 2 \pi(1- \cos \theta)$ is the solid angle subtended by the curve  $\Gamma$ at the centre of the sphere. For the pure neutrino states, we obtain the expression 
\begin{equation} \label{eq34}
    \gamma_G^{AA}=\pm \pi(1- \cos \theta),
\end{equation}
where positive and negative signs correspond to $\ket{\nu_e}$ and $\ket{\nu_\mu}$ respectively. The expression \eqref{eq34} is the Aharonov-Anandan phase obtained in Ref. \cite{Blasone:1999tq}. \\
$(iii)$ \textit{Neutrino propagation in non-dissipative matter}. In presence of a medium, the neutrino oscillation parameters are modified due to coherent forward scattering of the neutrinos with the background particles. If the medium has constant density, the modification is of the form : $\theta \rightarrow \theta_m$ and $\Delta m^2 \rightarrow \Delta m_m^2$, where $\theta_m$ and $\Delta m_m^2$ are mixing angle and mass-squared difference in the medium. The cyclic geometric phase \eqref{eq34} in this case becomes
\begin{equation} \label{eq35}
    \gamma_G^{AA}= \pm \pi\big(1- \cos \theta_m\big)= \pm \pi\Bigg[1- \frac{\cos \theta -V}{\sqrt{1- 2V \cos \theta + V^2}}\Bigg],
\end{equation}
where $V= 2 E V_{cc}/\Delta m^2$, $V_{cc}$ being the charged-current potential. Thus we obtain the result derived in Ref. \cite{Dajka:2011zz} for neutrino geometric phase in dissipation-less matter. 

\subsection{Geometric phase versus quantum coherence} 
\begin{figure}[t] 
\Centering \includegraphics[width=70mm]{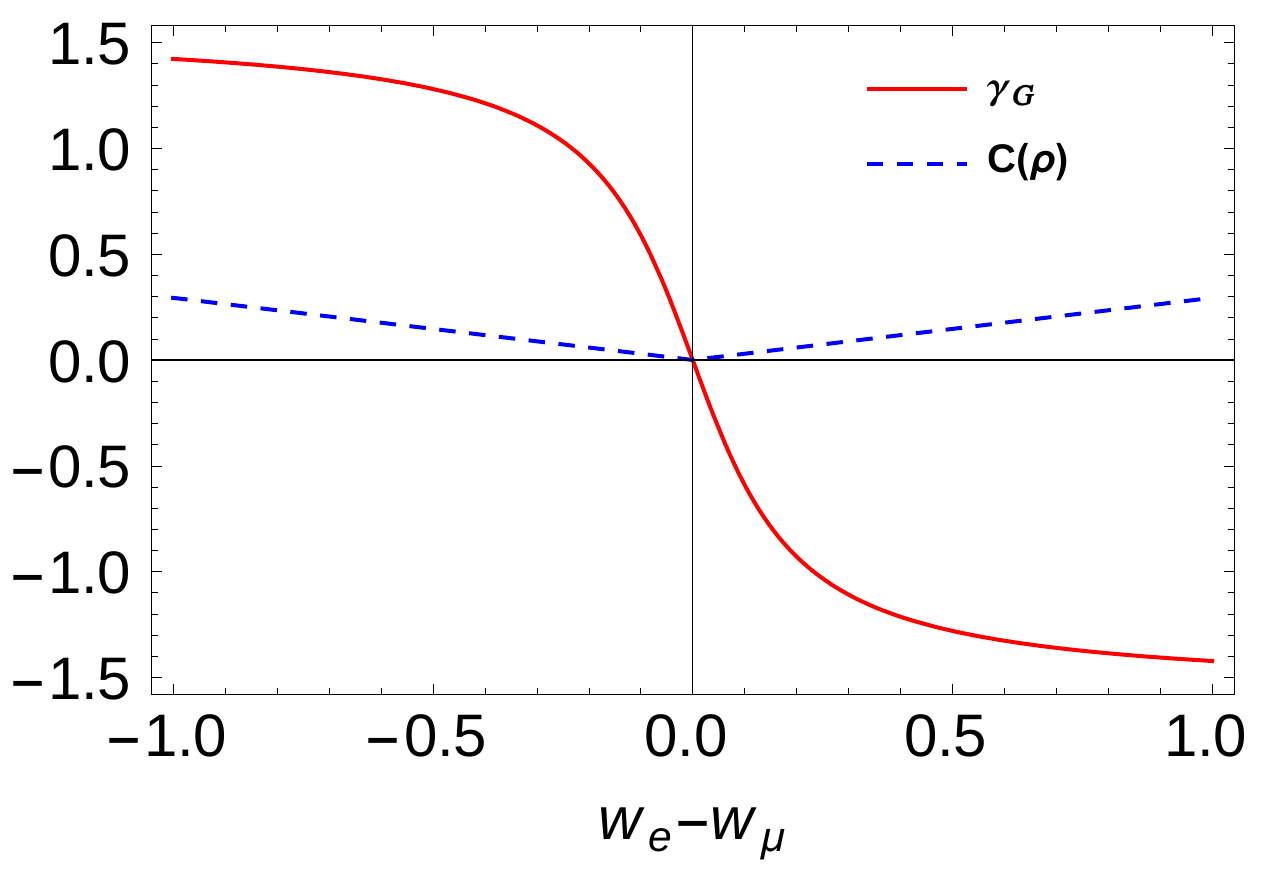}
\caption{Comparison of mixed state geometric phase \eqref{eq31} and quantum coherence \eqref{eq37} with respect to coherence parameter $w_e- w_\mu$. The neutrino oscillation parameters are taken as: $\Delta m^2= 2.5 \times 10^{-3}$ eV$^2$, $\theta= 48.6^\degree$, $L/E= 520$ (km/GeV). } 
\label{fig-1}
\end{figure}

The study of coherence properties of neutrino beams can offer useful insights about the neutrino propagation in a medium \cite{Bruss:1988fr, Jones:2014sfa}.
The form of \eqref{eq31} shows explicit dependence of geometric phase on the quantity $w_e- w_\mu$, which is the relative amount of $\nu_e$ and $\nu_\mu$ neutrinos present in the beam. For a maximally incoherent beam in which $w_e = w_\mu$, the geometric phase \eqref{eq31} vanishes.  Thus the expression \eqref{eq31} carries the information about the coherence content of the neutrino beam. Recently, quantum coherence in neutrino oscillations has been studied using tools from quantum information theory \cite{Song:2018bma}, wherein coherence is quantified using the $l_1-$norm:
\begin{equation} \label{eq36}
    C(\rho) =\frac{1}{d-1} \sum_{\substack{k, j \\ k \neq j}} |\rho_{k,j}|,
\end{equation}{}
where $\hat{\rho}$ is $d\times d$ representation of the density matrix of the system in a given basis. In our case, using the expression \eqref{eq19} for the density matrix, we have 
\begin{align} \label{eq37}
    C(\rho) =& 2|(w_e- w_\mu)||\psi_{ee} \psi_{e \mu}| \nonumber \\
    =& 2|(w_e- w_\mu)||\sin 2\theta \sin (\omega_p t/2)| \nonumber \\&|(1- \sin^2 2\theta \sin^2 (\omega_p t/2))^{1/2}|.
\end{align}
Comparing \eqref{eq37} with the expression of geometric phase \eqref{eq19}, we see that both the quantities are sensitive to the factor $w_e- w_\mu$, which defines the coherence content of the neutrino beam. In Figure \ref{fig-1} we plot the two quantities as a function of $w_e-w_\mu$ for typical oscillation parameters. It can be clearly seen that as the neutrino beam becomes more coherent, both geometric phase \eqref{eq31} and quantum coherence \eqref{eq37} reach their respective maximum values. Also, for completely incoherent beam they both vanish. Thus both quantities contain information about the \textit{quantumness} of the neutrino beam and can be considered as a measure of coherence for two-flavor neutrino oscillations.
\section{Three flavor oscillations}
Let us now consider the case of three flavor neutrino oscillations. In this case the geometric picture of neutrino oscillations involves precession of an eight dimensional polarization vector around a magnetic field \cite{Xing:2011zza}. The space of density operators corresponds to $SU(3)/(U(1) \times U(1))$, when the density matrix has non-degenerate eigenvalues \cite{byrd2007geometry}. Even though pictorial representation is too complicated to visualize for three flavor oscillations, most of the mathematical expressions admit a straightforward generalization of the results in the preceding section. 

The neutrino beam is described by initial density matrix 
\begin{equation} \label{eq38}
    \rho(0) = \mbox{Diag}(w_e, w_\mu, w_\tau). 
\end{equation}
The evolution of neutrino flavor states is governed by the unitary operator 
\begin{equation}
    \mathcal{U}(t) = \begin{pmatrix}
    \psi_{ee}(t) & \psi_{\mu e}(t)& \psi_{\tau e}(t) \\
    \psi_{e \mu}(t) & \psi_{\mu \mu}(t)& \psi_{\tau \mu}(t)\\
    \psi_{e \tau}(t) & \psi_{\mu \tau}(t)& \psi_{\tau \tau}(t)
    \end{pmatrix},
\end{equation}
where the transition amplitudes are given by 
\begin{equation} \label{eq40}
    \psi_{\alpha \beta}(t)= \sum_i U^{\ast}_{\alpha i} U_{\beta i} e^{-i E_i t},
\end{equation}
where $E_i = m_i^2/2E; i=1,2, 3$. For the mixing matrix $U$, we assume the standard Dirac parametrization with three mixing angles $\theta_{12}$, $\theta_{13}$, $\theta_{23}$ and a CP (charge-conjugation and parity)-violating phase $\delta$ (see Eq.(6.191) in \cite{Giunti:2007ry}).  
The time evolved density matrix $\rho(t)= \mathcal{U}(t)\rho(0)\mathcal{U}^{\dagger}(t)$ can now be written as
\begin{equation}
    \rho(t) = \sum_{\alpha= e, \mu, \tau} w_\alpha \begin{pmatrix}
    \psi_{\alpha e}(t) \\ \psi_{\alpha \mu}(t) \\ \psi_{\alpha \tau}(t)
    \end{pmatrix}
     \begin{pmatrix}
    \psi^{\ast}_{\alpha e}(t) & \psi^{\ast}_{\alpha \mu}(t) & \psi^{\ast}_{\alpha \tau}(t)
    \end{pmatrix},
\end{equation}
where $\ket{e_\alpha}= \big(\psi_{\alpha e}~\psi_{\alpha \mu}~ \psi_{\alpha \tau} \big)^T,~ \alpha= e, \mu, \tau$ are the eigenvectors of $\rho$. The geometric phase can  now be obtained from \eqref{eq9}:
\begin{align} \label{eq42}
    \gamma_G=& \arg \Bigg\{\sum_{\alpha= e, \mu, \tau} w_\alpha \psi_{\alpha \alpha}(\tau) \\& \quad \times \exp\Big(- \int_0^\tau dt \sum_{\beta= e, \mu, \tau} \psi_{\alpha \beta}^{\ast}(t) \Dot{\psi}_{\alpha \beta}(t) \Big)\Bigg\},
\end{align}
which is a simple generalization of \eqref{eq29}. However it is too complicated to write \eqref{eq42} in a form analogous to \eqref{eq31}. 
A relatively simpler expression can be obtained for pure states. Let us consider the geometric phase for $\ket{\nu_e}$, for which \eqref{eq42} reduce to 
\begin{equation} \label{eq43}
\gamma_G^P= \arg \Bigg\{\psi_{ee}(\tau) \exp\Big(- \int_0^\tau dt \sum_{\beta= e, \mu, \tau} \psi_{e \beta}^{\ast}(t) \Dot{\psi}_{e \beta}(t) \Big)\Bigg\}. 
\end{equation}
Substituting $\psi_{\alpha \beta}$ from \eqref{eq40}, we obtain the following expression:
\begin{align}\label{eq44}
\gamma_G^P =& \tan^{-1}\frac{\cos 2 \theta_{12} \cos^2\theta_{13} \sin{\xi \tau}- \sin^2 \theta_{13}\sin\big((2q-1)\xi \tau\big)}{\cos^2\theta_{13} \cos{\xi \tau}+ \sin^2 \theta_{13}\cos\big((2q-1)\xi \tau\big)} \nonumber \\& + \big(2 \sin ^2 \theta_{12} \cos^2 \theta_{13}+ 2 q \sin^2 \theta_{13}-1\big) \xi \tau,
\end{align}
where $\xi= (E_2-E_1)/2= \Delta m_{21}^2/4E$ and $q= (E_3-E_1)/(E_2-E_1)= \Delta m_{31}^2/\Delta m_{21}^2$. The above expression matches the pure state geometric phase for $\ket{\nu_e}$ derived in Ref. \cite{Wang:2000ep}. Note that \eqref{eq44} is independent of the CP-violating phase $\delta$. However, it can be shown that the pure state geometric phases for $\ket{\nu_\mu}$ and $\ket{\nu_\tau}$ include non-trivial dependence on $\delta$. 
\section{Conclusions}
Neutrino oscillations represent a phenomenon in which quantum mechanical effects are observed at long distance scales. This provides us the opportunity to study the quantum mechanical features of this system such as geometric phase and quantum coherence in a unique manner. In particular, the appearance of geometric phases in neutrino oscillations have been pointed out in several previous studies. However, all of them consider the case of pure neutrino states, which cannot be realized in a typical scenario.       

In this work, we consider the more general case of an incoherent beam of neutrinos, and derive the expressions for geometric phase in both two flavor and three flavor models. We discussed the geometry of the state space of neutrino oscillations and its connection with the appearance of a geometric phase. For two flavor oscillations, the geometric phase is shown to be independent of the Majorana phase, however for three flavor oscillations the geometric phase shows non-trivial dependence on the Dirac CP-violating phase. We also show that our results generalize the previously obtained expressions of the pure state geometric phase for neutrino oscillations in vacuum and in non-dissipative matter. In addition, the comparison between geometric phase and information-theoretic quantum coherence is also highlighted. 

\section*{Acknowledgments}
The author would like to thank Dr. S.R. Jain for useful comments on the manuscript. 

\bibliographystyle{elsarticle-num}
\bibliography{main.bbl}
\end{document}